\documentstyle[11pt]{article}
\newcommand{\be}{\begin{equation}}
\newcommand{\ee}{\end{equation}}
\newcommand{\abz}{\hspace*{.5in}}

\refstepcounter{equation}

\begin{document}
\date{}
\begin{titlepage}
\begin{flushright}
gr--qc/9312019
\end{flushright}
\begin{center}
{\large\bf Quantum Kinetic Equations and Cosmology}\footnote{submitted to
Phys. Lett. A}
\vskip 1cm
{\bf Oleg A. Fonarev}
\vskip 0.5cm
Racah Institute of Physics, The Hebrew University
\\Jerusalem 91904, Israel
\\E--mail: OLG@vms.huji.ac.il
\vskip 2cm
{\bf Abstract}
\end{center}
\begin{quote}
We analyse quantum--kinetic effects in the early Universe. We show that quantum
corrections to the Vlasov equation give rise to a dynamical variation of the
gravitational constant. The value of the gravitational constant at the Grand
Uni\-fication epoch is shown to differ from its present value to about $10^{-4}
\div 10^{-3} \% $.
\end{quote}
\thispagestyle{empty}
\setcounter{page}{0}
\end{titlepage}
\newpage
\setcounter{page}{1}
\abz
It is commonly believed that, after the inflation has finished, the Universe
evolves according to the standard cosmological scenario, where matter is viewed
as a relativistic gas of point particles characterized by a classical
distribution function (see, for instance, Ref. \cite{kn:padman}).
The evolution of the distribution function is assumed governed by a
relativistic
 Boltzmann equation \cite{kn:stewart,kn:israel}.
Two different approximations have been used: either particle collisions are
completely neglected -- this is valid when the expansion rate of the Universe,
${H}(t)$, is much higher than reaction rates, ${\Gamma_{i}}(t)$;
or there is collisional equilibrium (detailed balancing) -- this is valid in
the opposite limit, ${H}(t) < {\Gamma_{i}}(t)$. \\
\abz
The distribution function of collision--free particles in a Robertson--Walker
Universe is a function of the magnitude of the 3--momentum of a particle,
$|\vec{p}|$, multiplied by the expansion factor, ${a}(t)$ \cite{kn:ehlers}.
If a species was in thermal equilibrium at some time, its distribution function
after decoupling is given by the "frozen--in" form \cite{kn:padman}:
\be
{f_{0}}(p) = \frac{g}{(2\pi\hbar)^{3}} \left[ \exp \left\{ \sqrt{(\frac{{a}(t)
\vec{p}}{T_{0} {a}(t_{0})})^{2} + \frac{m^2}{T_{0}^{2}}} -
\frac{\mu_{0}}{T_{0}} \right\} \pm 1 \right]^{-1} \; , \label{eq:f0}
\ee
where $m$ is the mass of a particle, $g$ is the spin degeneracy factor, $T_{0}$
is the temperature at time $t_{0}$ when the decoupling occurs and $\mu_{0}$ is
the chemical potential at temperature $T_{0}$.
The upper sign (+1) corresponds to the Fermi--Dirac statistics and the lower
sign (-1) is for the Bose--Einstein statistics. \\
\abz
A natural question arises to what extent the classical picture is adequate and
how it is distorted by quantum effects, such as {\em Zitterbewegung}
\cite{kn:degroot,kn:habib}.
We address this question by analyzing quantum corrections in Friedmann models
caused by the interplay of curvature and non--locality of a quantum "particle".
\\
\abz
One of the most effective tools for the computing semiclassical quantum
corrections is a Wigner function method \cite{kn:wigner,kn:carruthers}.
It allows one to find quantum corrections to physical observables, once a
classical distribution function is known. \\
The notion of Wigner functions has been generalized to curved spacetime by
using different approaches, Refs.\ \cite{kn:winter}--\cite{kn:fonarev2}.
All the approaches are locally equivalent \cite{kn:fonarev2,kn:habib} and lead
to Wigner--type kinetic equations supplemented by generalized mass--shell
constraints.
The equations are written in terms of formal adiabatic expansions, with a
typical term \cite{kn:fonarev2}
\be
\left( \hbar^{k} \, R^{\alpha}_{\nu_{1} \beta \nu_{2} ; \nu_{3} \ldots \nu_{k}}
\, \frac{\partial^{k}}{\partial p_{\nu_{1}} \ldots \partial p_{\nu_{k}}}
\right)^{n} \;
{f}(x,p) \;. \label{eq:term}
\ee
Here ${f}(x,p)$ is the generalized Wigner function, $R^{\alpha}_{\nu \beta
\mu}$
 is the Riemann tensor and the semicolon signifies the covariant
differentiation. \\
In a radiation dominated Friedmann--Robertson--Walker Universe \cite{kn:padman}
with line element
\be
ds^{2} = dt^{2} - {a^{2}}(t) \, \gamma_{ij} \, dx^{i} \, dx^{j} \; ,
\label{eq:rw}
\ee
where $\gamma_{i j}$ is a metric of a 3--space of constant curvature, an
adiabatic parameter in the perturbative expansion of the quantum kinetic
equations is
\be
\delta = ( \hbar {H}(t) / {T}(t) )^{2} \; , \label{eq:delta}
\ee
if one treats the distribution function (\ref{eq:f0}) as a classical limit of
the Wigner function.
 Here ${H}(t) = \frac{\dot{a}}{a}$ -- is the Hubble "constant" and ${T}(t) =
T_{0}
\frac{{a}(t_{0})}{{a}(t)}$ -- is the local temperature. One can show that in
the radiation dominated Friedmann model \cite{kn:borner} $\delta$ is given by
\be
\delta = {g}(T) T^{2} / M_{pl}^{2} \; , \label{eq:deltafr}
\ee
where ${g}(T)$ is the number of degrees of freedom of those particles which are
still relativistic at temperature ${T}(t)$, $M_{pl} \simeq 10^{19}$ GeV -- is
the Planck energy.
At the GUT temperature, $T_{GUT} \simeq 10^{15}$ GeV, if ${g}(T_{GUT})
\simeq 10^{2} \div 10^{3}$, one gets $\delta \simeq 10^{-6} \div 10^{-5}$
(the uncertainty in the value of ${g}(T)$ reflects the uncertainty in our
knowledge of particle physics at high energies; though the Standard Model is
in excellent agreement with all current data, it is not clear yet what
fundamental symmetry is responsible for Grand Unification \cite{kn:borner}).
Thus our estimates show that the Wigner function approach, which utilizes an
adiabatic expansion, is safely valid below the GUT scale. \\
\abz
For the sake of comparison, let us also consider quantum corrections due to
the vacuum polarization effects \cite{kn:birdav}.
In the radiation dominated Friedmann model the first nontrivial quantum--vacuum
correction to the energy density, $\rho_{cl}$, is $\rho_{cl} \cdot
\delta_{vac}$, with $\delta_{vac}$ given by \cite{kn:parker}
\begin{eqnarray}
\delta_{vac} & = & \alpha t^{2}_{pl} {H^{2}}(t) \nonumber \\
             & = & \alpha {g}(T) T^{4} / M_{pl}^{4} \; .
\end{eqnarray}
Here $\alpha$ is a dimensionless parameter, usually of the order of 1.
At the GUT temperature, with ${g}(T_{GUT}) \simeq 10^{2} \div 10^{3}$, one gets
$\delta_{vac}
\simeq 10^{-14} \div 10^{-13}$.
Therefore, one can neglect the quantum--vacuum effects when computing the
quantum--kinetic corrections at the first order in terms of $\delta$,
(\ref{eq:delta}) (the second adiabatic order). \\
\abz
Let us proceed now to a more detailed analysis of the lowest order
quantum--kinetic corrections in a Friedmann--Robertson--Walker Universe.
We shall consider a spatially flat Universe for simplicity, since, after an
inflation era, the dynamics insignificantly depends on the value of the
spatial curvature \cite{kn:linde}. \\
The model is described by the Einstein equation \cite{kn:padman}:
\be
\frac{1}{8 \pi G} (R_{0}^{0} - \frac{1}{2} R) = \rho_{tot} \; , \label{eq:ein}
\ee
where $R_{0}^{0} - \frac{1}{2} R = 3 \frac{{\dot{a}}^{2}}{a^{2}}$ in our case,
$G$ is the Newtonian constant and $\rho_{tot} = T^{0}_{0}$ is the energy
density of matter involving quantum corrections.
Since we have in mind applications to the early Universe, we shall assume that
all particles are massless.
It seems plausible that around the GUT scale a state of complete thermodynamic
equilibrium existed \cite{kn:borner}.
Therefore, the particles at high temperatures are described by the distribution
function (\ref{eq:f0}), with $m = \mu_{0} = 0$ (in this limit the distribution
function fulfils detailed balancing) and $T_{0}$ be the GUT temperature.
The vanishing of the collision integral implies that the lowest order quantum
corrections to the Boltzmann equation can be derived from noninteracting fields
coupling only to gravity.
The spin--curvature interaction \cite{kn:fonarev1,kn:fonarev2} vanishes in a
Robertson--Walker Universe, due to the symmetry of the metric.
Hence different fields give similar contributions to $\rho_{tot}$.
The difference in the statistics (the sign in Eq.\ (\ref{eq:f0})) only results
in a numerical factor, which can be taken into account by a proper definition
of the function ${g}(T)$. \\
Thus consider a real scalar fields $\varphi$ obeying the equation:
\be
( \Box - \xi R ) \, \varphi = 0 \; , \label{eq:klein}
\ee
where $\Box$ is the Laplace--Beltrami operator defined on the background
spacetime, $R$ is the Ricci scalar and $\xi$ is the nonminimal gravitational
coupling constant, $\xi = 1/6$ corresponds to conformal coupling
\cite{kn:birdav}.
Given the field equation, (\ref{eq:klein}), one is able to derive a quantum
corrected Vlasov equation which the  generalized Wigner function obeys
\cite{kn:winter,kn:hu,kn:fonarev2}. \\
The lowest order quantum corrections  to isotropic distributions in Robertson--
Walker spacetimes were found in \cite{kn:pirk} where the connection of the
result to the standard WKB--calculations is also discussed.
In \cite{kn:habib} a general analysis of the two alternative approaches to
statistical quantum field theory in curved spacetime is represented. \\
\abz
Let us outline a different from \cite{kn:pirk} derivation of the quantum
corrections, most suitable in a spatially flat Robertson--Walker spacetime (see
 our next paper \cite{kn:fonarevunp} for more de\-tails).
It is well known \cite{kn:birdav} that the scaled field $\overline{\varphi} =
{a}(t) \varphi$ obeys the Klein--Gordon equation in Minkowski spacetime with
metric $\eta^{\alpha \beta}$, in the presence of the potential ${V}(t) =
( \frac{1}{6} - \xi ) {a^{2}}(t) {R}(t)$:
\be
( \eta^{\alpha \beta} \partial_{\alpha} \partial_{\beta} + V ) \,
\overline{\varphi} = 0 \; . \label{eq:flatklein}
\ee
One can now define a Wigner function, ${\overline{f}}(x,p)$, for the field
$\overline{\varphi}$ in the Minkowski spacetime, derive a Wigner--type kinetic
equation, find a solution to it and  all the observables expressible in terms
of the Wigner function, and then perform a conformal transformation back to the
 Robertson--Walker spacetime.
A classical distribution function is invariant under conformal transformations
\cite{kn:fonarevunp}.
In our particular case, this can be seen from Eq.\ (\ref{eq:f0}):
$({a}(t) \vec{p})^{2} = p_{x}^2 + p_{y}^{2} + p_{z}^{2} \equiv
\vec{p}_{M}^{\,2}$ --
is the 3--momentum squared of a particle in the Minkowski spacetime (note that
the classical particles are massless because the potential ${V}(t)$ is of the
second adiabatic order).
The field equation (\ref{eq:flatklein}) yields the following set of equations
for the Wigner function ${\overline{f}}(x,p)$ \cite{kn:hu,kn:fonarev2}:
\be
( \hbar^{2} V - \eta^{\alpha \beta} p_{\alpha} p_{\beta} ) \, \overline{f} +
\frac{\hbar^{2}}{4} \eta^{\alpha \beta} \partial_{\alpha} \partial_{\beta}
\overline{f} = 0  \label{eq:kin}
\ee
\be
( \eta^{\alpha \beta} p_{\alpha} \partial_{\beta} + \frac{\hbar^{2}}{2} V_{,
\alpha} \frac{\partial}{\partial p_{\alpha}} ) \, \overline{f} = 0 \; .
\label{eq:mass}
\ee
Here $\partial_{\alpha} = \frac{\partial}{\partial x_{\alpha}}$ -- is the
partial derivative operator and $V_{, \alpha} = \partial_{\alpha} V$.
In the case when $V$ is a function of time only, Eqs.\ (\ref{eq:kin}) and
(\ref{eq:mass}) can be easily solved for isotropic distributions.
Up to the next adiabatic order, any function $\overline{f} = {F_{0}}(\vec{p}_
{M}^{\,2})\, {\delta}( \frac{1}{2} \hbar^2 V - \frac{1}{2} \eta^{\alpha \beta}
p_{\alpha} p_{\beta} )$, where $\delta$ is the Dirac $\delta$--function,
fulfils both (\ref{eq:kin}) and (\ref{eq:mass}). \\
The number--flux vector $\overline{N}_{\alpha}$ and the stress--energy tensor
$\overline{T}_{\alpha}^{\beta}$ for such distributions are easily found
\cite{kn:fonarevunp}:
\be
\overline{N}_{\alpha} = n_{2} U_{\alpha} \label{kn:number}
\ee
\be
\overline{T}_{\alpha}^{\beta} = \frac{1}{3} n_{3} (4 U_{\alpha} U^{\beta} -
\delta_{\alpha}^{\beta} ) + \frac{1}{6} \hbar^{2} V n_{1} ( 2 U_{\alpha}
U^{\beta} + \delta_{\alpha}^{\beta} )  \; . \label{eq:energymom}
\ee
Here $U_{\alpha} = \delta_{\alpha}^{0}, U^{\alpha} = \delta_{0}^{\alpha}$ and
\be
n_{k} = 4 \pi \, \int_{0}^{\infty} dp \, p^{k} \, {F_{0}}(p^{2}) \; .
\label{eq:nk}
\ee
Now one can compute the number--flux vector, $N_{\alpha}$, and the stress--
energy tensor, $T_{\alpha}^{\beta}$, in the original Robertson--Walker
spacetime, by applying the conformal transformation \cite{kn:fonarevunp}:
\be
N_{\alpha} = a^{-2} \, \overline{N}_{\alpha} \label{eq:numberrw}
\ee
\be
T_{\alpha}^{\beta} = a^{-4}\, \overline{T}_{\alpha}^{\beta} + \hbar^{2} (\xi -
\frac{1}{6}) n_{1} a^{-2} (R^{0}_{0} \delta_{\alpha}^{\beta} - 2 R_{0}^{0}
U_{\alpha} U^{\beta} + R U_{\alpha} U^{\beta}) \; . \label{eq:energymomrw}
\ee
Here $R_{0}^{0}$ is the $^{\circ}_{\circ}$--component of the Ricci tensor.
If one uses (\ref{eq:energymom}) in (\ref{eq:energymomrw}) and recalls that
$V = (\frac{1}{6} - \xi) a^2 R$, one can get the expression for the energy
density $\rho = T^{0}_{0}$ of the scalar field $\varphi$:
\be
\rho = a^{-4} n_{3} - \hbar^{2} (\xi - \frac{1}{6}) n_{1} a^{-2} (R_{0}^{0} -
\frac{1}{2} R) \; . \label{eq:energy}
\ee
The first term in (\ref{eq:energy}) is the classical energy density of
ultra--relativistic particles, whereas the second term represents the lowest
order quantum corrections.
The expression for the number--flux vector, (\ref{eq:numberrw}), indicates that
the density of particles, $n = n_{2} a^{-3}$, is not modified by the lowest
order quantum corrections.
It should be noted that, in general, function $F_{0}$ may implicitly involve
the Planck constant $\hbar$.
A different choice of $F_{0}$ corresponds to a different normalization of a
state vector \cite{kn:pirk}.
One can assign a one--particle Hilbert space such that an observer counts
particles as if they are classical. \\
Then, the structure of the quantum corrections in (\ref{eq:energy}) implies
that one can rewrite the Einstein equation (\ref{eq:ein}) as follows:
\be
\frac{3}{8 \pi {G_{eff}}(a)} \frac{\dot{a}^{2}}{a^{2}} = \rho_{cl} \; .
\label{eq:einnew}
\ee
Here $\rho_{cl}$ is the classical energy density, and the $G_{eff}$ stands for
the effective gravitational "constant" as measured by a local observer:
\be
G_{eff}^{-1} = G^{-1} + 8 \pi \hbar^{2} (\xi - \frac{1}{6}) n_{1} a^{-2} \; .
\label{eq:geff}
\ee
If the particles are described by the Planck distribution then Eq.\
(\ref{eq:geff}) gives:
\be
\frac{G}{{G_{eff}}(T)} = 1 + \frac{2}{3} \pi (\xi - \frac{1}{6}) \delta \; ,
\label{eq:ratio}
\ee
where $\delta$ is given in (\ref{eq:deltafr}). \\
Thus, for the minimal coupling ($\xi = 0$), the value of the gravitational
constant at the GUT temperature was greater than its present value to $10^{-4}
\div 10^{-3} \%$ (the quantum--kinetic cor\-rections are, of course, negligible
in the present Universe; about quantum effects in the cosmic microwave
background radiation, see also Ref.\ \cite{kn:messer}).  \\
\abz
We have shown that quantum effects may play an important role in the early
Universe.
Though our analysis did not cover inflation epochs, it seems likely that the
quantum--kinetic corrections can not be neglected during the inflation.
Unfortunately, the particle physics at energies above $10^{16}$ GeV is very
uncertain.
Moreover, estimates using the concept of asymptotic freedom indicate that at
temperatures $T > 10^{16}$ GeV thermodynamic equilibrium was not established
\cite{kn:borner}, that is the Universe is expected to be in a highly coherent
state.
The adiabatic expansions fail for such states and nonperturbative methods in
quantum kinetic theory in curved spacetime should be developed.
\vspace*{\baselineskip} \\
I would like to thank Professors L.\ P.\ Horwitz and J.\ D.\ Bekenstein for
helpful discussions.

\end{document}